\documentclass[
superscriptaddress,
twocolumn,
amsmath,amssymb,
aps,
prx,
nofootinbib,
]{revtex4-2}

\usepackage{comment}
\usepackage{tabularx}
\usepackage{graphicx}
\usepackage{dcolumn}
\usepackage{bm}
\usepackage{mathtools}
\usepackage{color}

\usepackage{hyperref}
\hypersetup{
	colorlinks=true,
	linkcolor=black,
	citecolor=black,
	urlcolor=black
}

\begin{document}

\title{An optical tweezer array of ultracold polyatomic molecules}

\author{Nathaniel B. Vilas}
\email{vilas@g.harvard.edu}
\affiliation{Department of Physics, Harvard University, Cambridge, MA 02138, USA}
\affiliation{Harvard-MIT Center for Ultracold Atoms, Cambridge, MA 02138, USA}

\author{Paige Robichaud}
\affiliation{Department of Physics, Harvard University, Cambridge, MA 02138, USA}
\affiliation{Harvard-MIT Center for Ultracold Atoms, Cambridge, MA 02138, USA}

\author{Christian Hallas}
\affiliation{Department of Physics, Harvard University, Cambridge, MA 02138, USA}
\affiliation{Harvard-MIT Center for Ultracold Atoms, Cambridge, MA 02138, USA}

\author{Grace K. Li}
\affiliation{Department of Physics, Harvard University, Cambridge, MA 02138, USA}
\affiliation{Harvard-MIT Center for Ultracold Atoms, Cambridge, MA 02138, USA}

\author{Lo\"ic Anderegg}
\affiliation{Department of Physics, Harvard University, Cambridge, MA 02138, USA}
\affiliation{Harvard-MIT Center for Ultracold Atoms, Cambridge, MA 02138, USA}

\author{John M. Doyle}
\affiliation{Department of Physics, Harvard University, Cambridge, MA 02138, USA}
\affiliation{Harvard-MIT Center for Ultracold Atoms, Cambridge, MA 02138, USA}

\date{\today}

\begin{abstract}

Polyatomic molecules have rich structural features that make them uniquely suited to applications in quantum information science~\cite{yu2019scalable,wei2011entanglement,tesch2002quantum}, quantum simulation~\cite{wall2013simulating,wall2014quantum,wall2015realizing}, ultracold chemistry~\cite{heazlewood2021towards}, and searches for physics beyond the Standard Model~\cite{kozyryev2017precision,kozyryev2021enhanced,hutzler2020polyatomic}. However, a key challenge is fully controlling both the internal quantum state and the motional degrees of freedom of the molecules. Here, we demonstrate the creation of an optical tweezer array of individual polyatomic molecules, CaOH, with quantum control of their internal quantum state. The complex quantum structure of CaOH results in a non-trivial dependence of the molecules' behavior on the tweezer light wavelength. We control this interaction and directly and nondestructively image individual molecules in the tweezer array with $>$90\% fidelity. The molecules are manipulated at the single internal quantum state level, thus demonstrating coherent state control in a tweezer array. The platform demonstrated here will enable a variety of experiments using individual polyatomic molecules with arbitrary spatial arrangement.

\end{abstract}

\maketitle

Ultracold molecules are an important frontier in quantum science due to their rich internal structure, large intrinsic electric dipole moments, and as a bridge between ultracold physics and quantum chemistry. Identified scientific applications include quantum simulation~\cite{micheli2006toolbox,gorshkov2011quantum,wall2014quantum}, quantum information processing~\cite{demille2002quantum,yelin2006schemes,ni2018dipolar,sawant2020ultracold,tesch2002quantum,wei2011entanglement,yu2019scalable,albert2020robust}, quantum chemistry~\cite{heazlewood2021towards}, collisional physics~\cite{cheuk_2020}, quantum metrology and clocks~\cite{kondov2019molecular}, and precision searches for physics beyond the Standard Model~\cite{kozyryev2017precision,kozyryev2021enhanced,norrgard2019nuclear}. One promising platform for realizing many of these goals is an optical tweezer array, which can allow for positioning of individually trapped particles in arbitrary geometries, as well as dynamic rearrangement of their positions~\cite{endres2016atom,barredo2018synthetic,bluvstein2022quantum}. Optical tweezer arrays have already proven to be extremely successful for ultracold atoms~\cite{kaufman2021quantum,browaeys2020many}, enabling the realization of high-fidelity quantum processors~\cite{levine2019parallel,bluvstein2022quantum,evered2023high} and quantum simulators~\cite{bernien2017probing,ebadi2021quantum,scholl2021quantum} based on Rydberg interactions, as well as the development of tweezer-based optical clocks for precision metrology~\cite{norcia2019seconds, madjarov2019atomic, young2020half} and the study of ultracold collisions~\cite{reynolds2020direct}. Tweezer arrays of diatomic molecules have also been achieved~\cite{anderegg2019optical,zhang2022optical,ruttley2023formation}, as has the observation of dipolar interactions and two-molecule entangling gates~\cite{holland2023on,bao2023dipolar}. Recently, Raman sideband cooling of these molecules to the motional ground state of the trap has demonstrated full quantum control of both internal and external states of laser cooled diatomic molecules~\cite{bao2023raman,lu2023raman}.

Polyatomic molecules, as compared to atoms and diatomic molecules, have non-trivial additional degrees of freedom that can be harnessed for quantum science and technology and precision measurement applications~\cite{augenbraun2023review}. For example, polyatomic molecules generically contain very closely spaced opposite-parity states that allow the molecule to be polarized in the laboratory frame using small applied electric fields, with a structure that is ideal for many applications~\cite{kozyryev2017precision,yu2019scalable,hutzler2020polyatomic}. Such states allow quantum information processing schemes with switchable dipole-dipole interactions~\cite{yu2019scalable,wei2011entanglement}, as well as natural simulation of quantum magnetism models~\cite{wall2013simulating,wall2014quantum,wall2015realizing}. The large number of internal states in polyatomic molecules could also be useful for encoding error-corrected qubit states~\cite{albert2020robust}, or for creating qudit~\cite{sawant2020ultracold} states enabling many bits of quantum information to be contained in a single physical molecule.
Additionally, polyatomic molecules are promising for searches for physics beyond the Standard Model (BSM), as they contain states with high BSM sensitivity and strong isolation from background noise and decoherence sources~\cite{kozyryev2017precision,anderegg2023quantum,kozyryev2021enhanced,takahashi2023engineering}.
All of these applications could either be enabled by, or benefit from, isolating and controlling individual polyatomic molecules in optical tweezer arrays. Towards these scientific goals, polyatomic molecules have been laser cooled~\cite{augenbraun2023review,kozyryev2017sisyphus,vilas2022magneto} and trapped in electrostatic~\cite{zeppenfeld2012sisyphus,prehn2016optoelectrical}, magnetic~\cite{liu2017magnetic}, and optical~\cite{hallas2023optical,anderegg2023quantum} traps, but they have not previously been controlled at the single particle level.

In this work, we demonstrate trapping of polyatomic molecules in an array of optical tweezers. Laser-cooled CaOH molecules are trapped in a magneto-optical trap (MOT) and transferred into a large optical dipole trap (ODT), and then single molecules are loaded into optical tweezers. During loading, a collisional blockade occurs~\cite{schlosser2001sub}, ensuring that each tweezer contains at most one molecule. The molecules are non-destructively imaged in the tweezers, enabling identification of loaded tweezers and re-imaging later in the experimental sequence. We observe that the loading and imaging efficiencies depend strongly on the tweezer light wavelength due to the rich spectrum of excited states in CaOH. We characterize this dependence and determine optimal parameters to control these effects. Finally, we prepare individually trapped CaOH molecules in a single quantum state within the lowest lying parity-doubled vibrational bending mode, an identified resource for many quantum science goals.

\section*{Experimental Overview}

The experiment begins with $\sim$$10^4$ laser-cooled CaOH molecules loaded in a magneto-optical trap (MOT) at $\sim$1~mK~\cite{vilas2022magneto}.
In order to achieve sufficient density for tweezer loading~\cite{anderegg2019optical}, the molecules are then cooled into an optical dipole trap (ODT) ($\sim$600~$\mu$K trap depth, 25~$\mu$m waist, 1064~nm light). A combination of $\Lambda$-enhanced grey molasses cooling and single-frequency (SF) cooling~\cite{hallas2023optical} loads the ODT, as well as six optical tweezer traps arrayed inside the ODT.

The tweezers are formed from a 784.5~nm laser beam that is diffracted through an acousto-optic deflector driven with multiple rf tones to generate the tweezers. The tweezers are projected using a 0.4 NA aspheric lens mounted inside the vacuum chamber, resulting in a beam waist of $\sim$2~$\mu$m for each trap. For this work, we use a static spacing of 11~$\mu$m between tweezers and a tweezer trap depth of 1.6~mK during loading, corresponding to approximately 150~mW of optical power per trap. To improve tweezer loading success and homogeneity across the array, we raster the ODT position along the array axis at a frequency of 100~Hz during loading.

After 80~ms of tweezer loading, the ODT and the SF cooling light are turned off, allowing molecules that are not in a tweezer to fall away. A 4~ms pulse of $\Lambda$-imaging light~\cite{cheuk_2020, hallas2023optical} is then applied to induce light-assisted collisions in tweezers with more than one molecule loaded, ensuring collisional blockade, wherein each tweezer contains at most one single molecule~\cite{schlosser2001sub}. After loading, the tweezer trap depth is lowered to 1.4~mK for optimal imaging fidelity.

\begin{figure}
    \centering
    \includegraphics{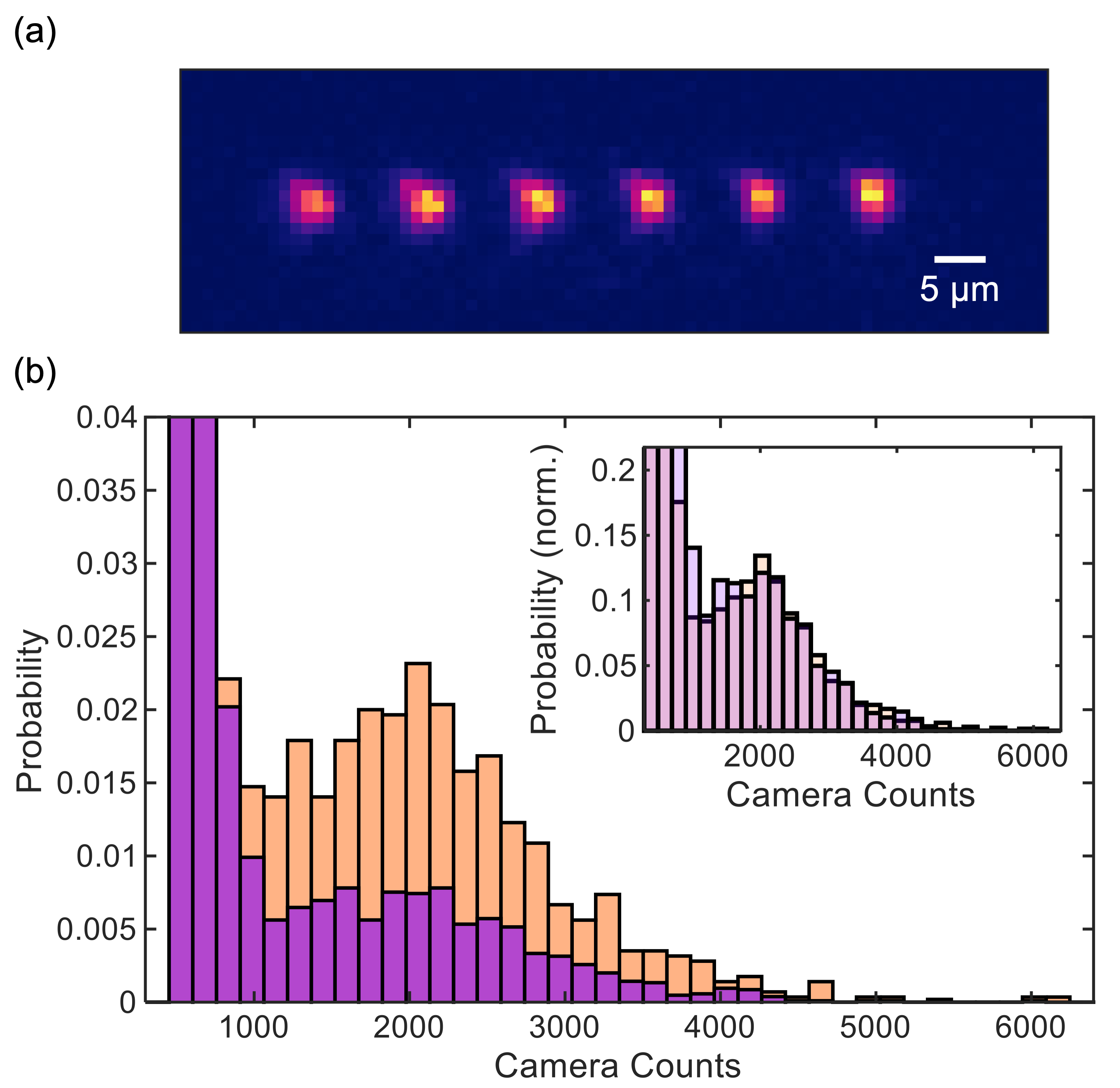}
    \caption{\textbf{An optical tweezer array of CaOH molecules.} (a) Averaged image of the CaOH tweezer array, attained by imaging the molecules for a 50~ms duration and averaging over hundreds of iterations of the experimental sequence. (b) Histograms of collected fluorescence for 15~ms duration tweezer images at average loading probabilities of 31\% (orange) and 13\% (purple). Inset: histograms normalized by loading rate, indicating that the shape of the loaded molecule peak does not change with the loading probability.}
    \label{fig:1}
\end{figure}

Molecules in the tweezers are imaged using $\Lambda$-enhanced grey molasses cooling, which allows thousands of photons to be scattered~\cite{cheuk2018lambda, hallas2023optical}. Fluorescence from the $\widetilde{A}^2\Pi_{1/2}(000) \rightarrow \widetilde{X}^2\Sigma^+(000)$ cooling transition at 626~nm is collected via the in-vacuum asphere, and then imaged onto an EMCCD camera. We find that optimal imaging occurs only at specific tweezer wavelengths and requires several of the vibrational repumping lasers used for imaging to be tuned by $\sim$100~MHz from their free-space values, as discussed in detail below. Fig. \ref{fig:1}a shows an averaged image of the tweezer array.

Histogram analysis of the intensity of fluorescence from the molecules reveals single molecule trapping in the tweezers.
Fig. \ref{fig:1}b shows histograms of camera counts attained from fluorescence imaging of the molecules in tweezers. Histograms are shown for 15~ms images and for two average tweezer loading probabilities of 31\% (orange) and 13\% (purple), tuned by adjusting the spatial overlap of the ODT with the tweezer array. The narrow peak on the left corresponds to empty tweezers, while the broader peak on the right corresponds to loaded tweezers; choosing a threshold value between the two peaks enables classification of single shot images as corresponding to loaded or empty tweezers. If multiple molecules were loaded in each tweezer, the average brightness would decrease as the loading fraction decreased, causing the right hand peak of the histogram to shift to the left. In contrast, as shown in the inset of Fig.~\ref{fig:1}b, rescaling the histograms by the measured loading rate shows that the shapes of the two histograms are identical, indicating that only single molecules are loaded.

\section*{Excited Electronic States}

\begin{figure}
    \centering
    \includegraphics{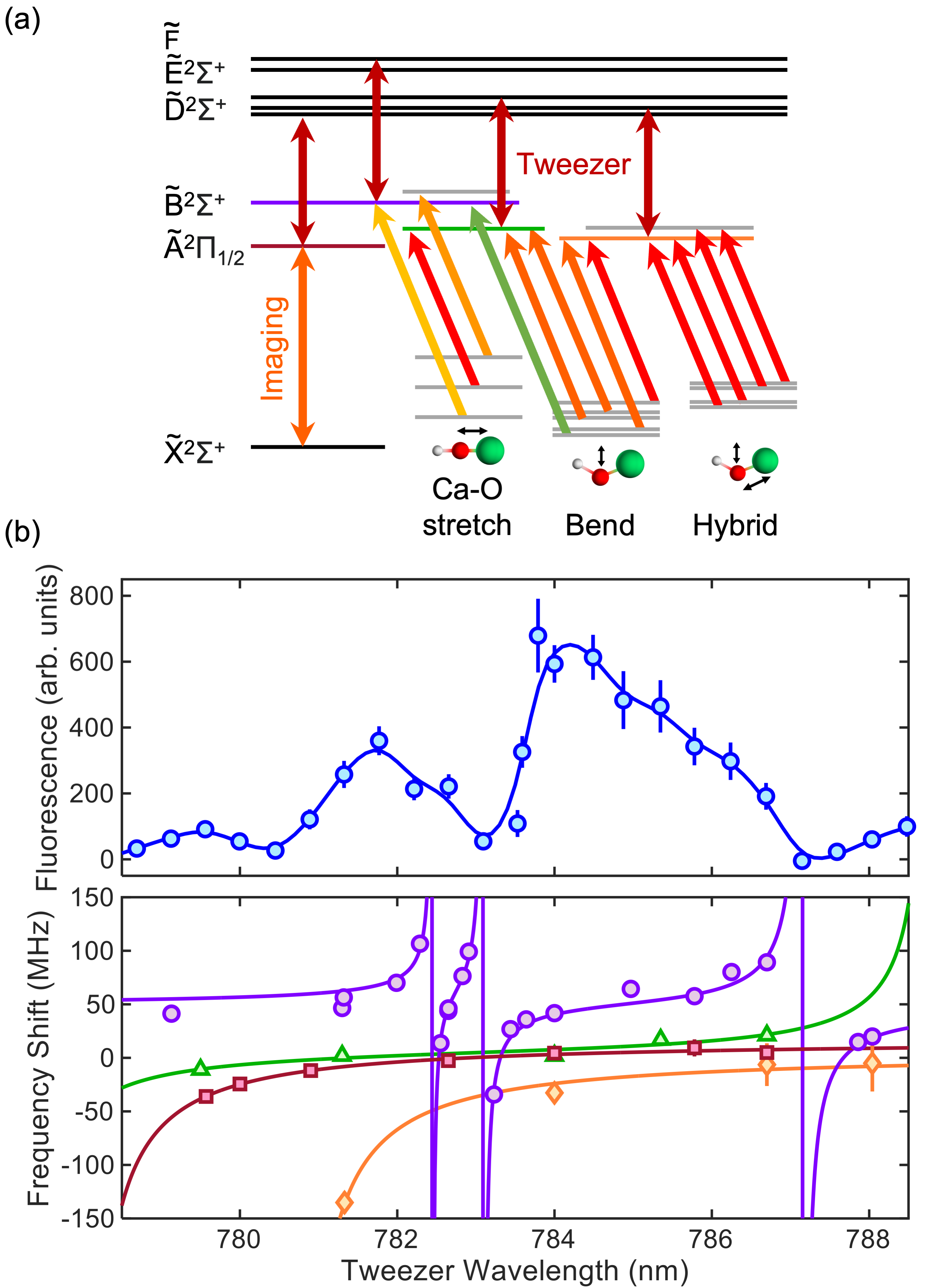}
    \caption{\textbf{Wavelength dependence of CaOH trapping.} (a) Level diagram showing the cooling and repumping lasers used for $\Lambda$ imaging (arrows connecting $\widetilde{X}^2\Sigma^+$ to the $\widetilde{A}^2\Pi_{1/2}$ and $\widetilde{B}^2\Sigma^+$ manifolds), as well as the high-lying electronic states to which the $\sim$780~nm tweezer light (dark red arrows) couples. (b) Top plot: average tweezer fluorescence vs. wavelength for 50~ms array images. The solid curve is a guide to the eye. Bottom plot: measured relative AC Stark shifts of the $\widetilde{A}^2\Pi_{1/2}(000)$ (red squares), $\widetilde{B}^2\Sigma^+(000)$ (purple circles), $\widetilde{A}^2\Pi_{1/2}(100)$ (green triangles), and $\widetilde{A}(010)\kappa^2\Sigma^{(-)}$ (orange diamonds) states. Solid curves are fits used to extract line positions and coupling strengths~\cite{Supplemental}. Error bars represent 68\% confidence intervals.}
    \label{fig:2}
\end{figure}

A general challenge of optically manipulating and trapping polyatomic molecules is the relatively large density of excited states compared to atoms and diatomic molecules. This shows up specifically in our experiment as molecular loss dependent on the wavelength of the tweezer light. We characterize the manifold of excited states near the tweezer wavelength, model the interactions of the molecules with the tweezer light, and are then able to tune the wavelength of the tweezer light to control loss.

Energy levels relevant to molecular imaging are shown in Fig.~\ref{fig:2}a. The high-lying electronic potentials $\widetilde{D}^2\Sigma^+$, $\widetilde{E}^2\Sigma^+$, and $\widetilde{F}$~\cite{pereira1996observation} are coupled to the $\widetilde{A}^2\Pi_{1/2}$ and $\widetilde{B}^2\Sigma^+$ states by the tweezer light. This has the effect of (a) shifting cooling and repumping imaging lasers out of resonance due to AC Stark shifts and (b) sometimes directly exciting molecules to these electronic states, which can lead to loss from imaging excitation.

Fig. \ref{fig:2}b (top plot) plots the overall brightness of molecules in the tweezers (a combination of loading fraction and imaging efficiency) as a function of tweezer light wavelength over the range 779-788~nm. This was done using fixed imaging and repumping frequencies chosen to optimize the signal at 784~nm. While bright tweezers are observable around 784~nm and 782~nm, they are very dim in several wavelength regions, notably around 783~nm, above 787~nm, and below 780~nm.

To characterize the wavelength dependence, we measure the AC Stark shifts of the laser cooling excited states as a function of tweezer wavelength (Fig. \ref{fig:2}b, bottom plot). Specifically, we study the $\widetilde{A}(000)$, $\widetilde{B}(000)$, $\widetilde{A}(100)$, and $\widetilde{A}(010)$ states, which are the most important excited states used for CaOH optical cycling~\cite{vilas2022magneto}. At each tweezer wavelength, we scan the frequency of a cooling or repumping laser addressing the target electronic state in order to maximize the tweezer brightness during imaging~\cite{Supplemental}. The offset between the optimal frequency and the known free-space resonance is the AC Stark shift. Because the tweezer is far detuned from all transitions out of the $\widetilde{X}$ manifold, the wavelength dependence is assumed to arise from Stark shifts of the excited states only, while the ground state AC Stark shifts contribute a constant offset.

We fit the observed Stark shifts to analytic lineshapes to determine the position and strength of the excited state couplings~\cite{Supplemental}. The energy of the $\widetilde{A}(000)$ excited state exhibits a resonance at $777.9(6)$~nm, which is consistent with the expected wavelength for coupling from $\widetilde{A}^2\Pi_{1/2}(000)\rightarrow \widetilde{D}^2\Sigma^+(100)$~\cite{pereira1996observation}. The other $\widetilde{A}^2\Pi_{1/2}$ state vibrational levels exhibit similar behavior, due to coupling with vibrational levels of the $\widetilde{D}^2\Sigma^+$ electronic potential~\cite{Supplemental}.

While the resonances observed in the $\widetilde{A}^2\Pi_{1/2}$ excited states were expected based on previous state assignments in the literature~\cite{pereira1996observation}, the $\widetilde{B}^2\Sigma^+(000)$ state is shifted by high-lying states with no known literature assignment. Three features are observed in our data, at $782.4(1)$~nm, $783.1(1)$~nm, and $787.2(1)$~nm, corresponding to unassigned excited states in CaOH at 30804(2)~cm$^{-1}$, 30792(2)~cm$^{-1}$, and 30726(2)~cm$^{-1}$, respectively. These states lie slightly above known vibrational levels of the $\widetilde{E}^2\Sigma^+$ and $\widetilde{F}$ electronic states~\cite{pereira1996observation} and therefore likely belong to one or both of those electronic manifolds. Based on this data, we choose an optimal tweezer wavelength of 784.5~nm, which is well separated from the observed resonances and results in minimal loss.

\section*{Imaging of Molecules in the Tweezers}

\begin{figure}
    \centering
    \includegraphics{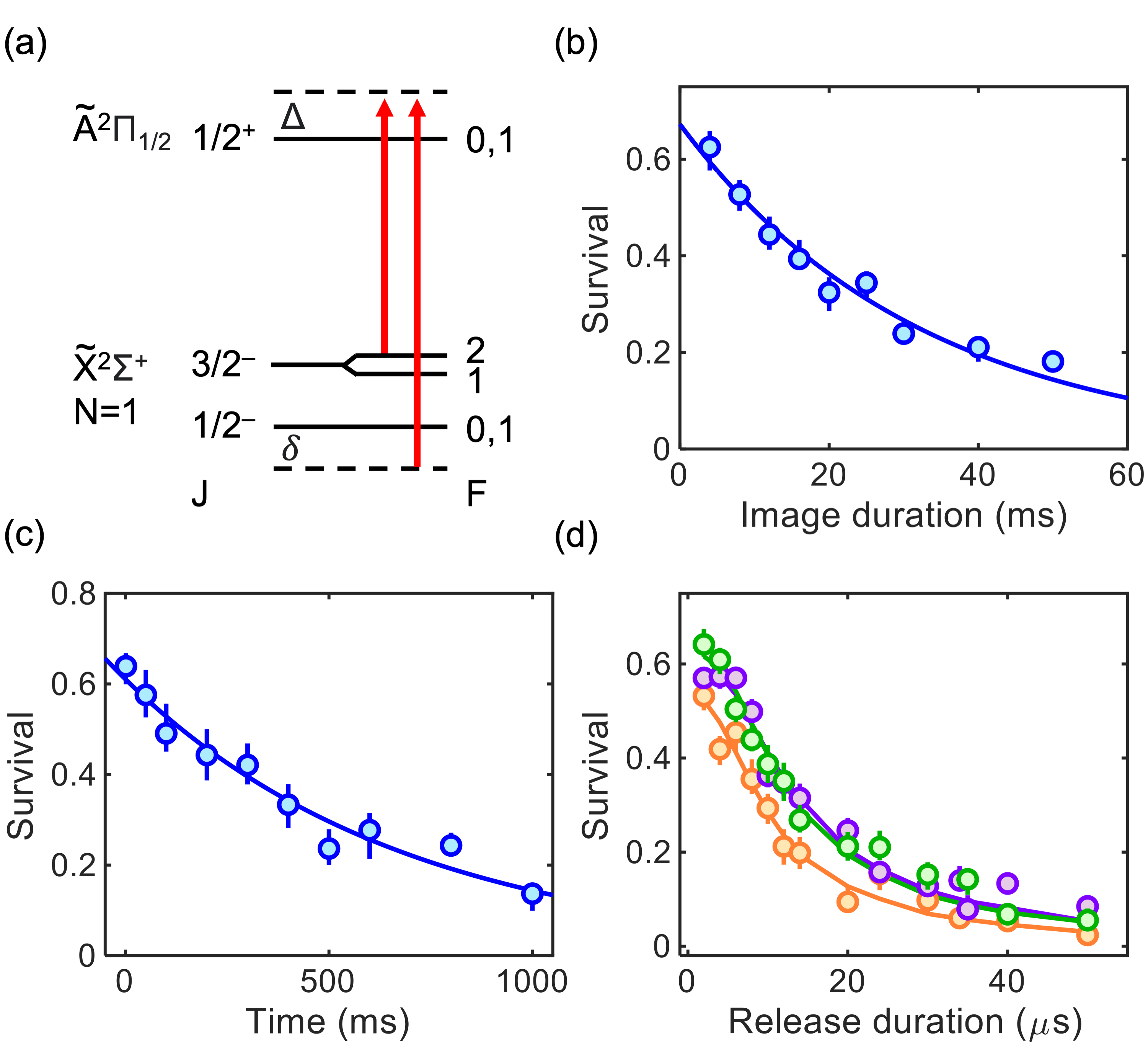}
    \caption{\textbf{Tweezer array imaging and characterization.} (a) Laser configuration used for $\Lambda$ imaging of the tweezer array, where $\Delta$ is the single-photon detuning and $\delta$ is the two-photon detuning. (b) Lifetime of CaOH molecules during imaging, measured to be 32(3) ms. (c) Tweezer hold lifetime, measured to be 690(70) ms, limited predominantly by background gas and blackbody excitation~\cite{hallas2023optical,vilas2023blackbody}. (d) Release-recapture temperature measurements for single-photon detunings of $\Delta = 16$~MHz (orange), 22~MHz (purple), and 28~MHz (green). Solid curves show the Monte Carlo fit results, indicating ratios of trap depth to temperature $\eta = 6(2)$, $12(1)$, and $9(1)$, respectively. Error bars represent 68\% confidence intervals.}
    \label{fig:3}
\end{figure}

We image CaOH molecules in the tweezer array using grey molasses cooling light in a $\Lambda$ configuration~\cite{cheuk2018lambda,hallas2023optical}, as shown in Fig. \ref{fig:3}a. The imaging parameters are chosen to maximize the number of photons that can be scattered from each tweezer-trapped molecule~\cite{Supplemental}. While the MOT and the tweezer loading are realized with the repumping lasers at their free-space resonance, between tweezer loading and imaging several repumping laser frequencies are tuned to the tweezer light-shifted resonances~\cite{Supplemental}. This ensures that the repumping lasers do not limit the imaging scattering rate and significantly increases the observed signal. Fluorescence histograms collected using a 15~ms imaging duration are shown in Fig. \ref{fig:1}b.

The imaging lifetime, $\tau$, is the average time the molecules survive during imaging before being lost from the tweezer or falling into a molecular dark state. To measure $\tau$ and the photon scattering rate of the molecules, $\Gamma$, we image for a variable duration, then wait 50~ms and image a second time (15~ms duration). The molecule survival (defined as the conditional probability that a molecule is detected in the second image, given that it was present in the first image) is plotted in Fig. \ref{fig:3}b as a function of the first image duration. The data are fit to an exponential decay with a $1/e$ imaging lifetime of $\tau = 32(3)$~ms. We also find for long imaging times $t\gg \tau$ that each molecule scatters $N_\text{im} = 8(1)\times 10^3$ photons, on average, before going dark to the imaging light~\cite{Supplemental}. Combining these measurements, we determine that the photon scattering rate is $\Gamma = 250(40)\times10^{3}$~s$^{-1}$ during imaging.
We measure the molecule lifetime in the absence of imaging light to be 690(70)~ms, which is determined predominantly by imperfect vacuum and by blackbody excitation (Fig.~\ref{fig:3}c)~\cite{hallas2023optical, vilas2023blackbody}.

The average number of photons scattered during long imaging pulses, $N_\text{im}$, is slightly less than the expected limit of approximately $1.6\times 10^4$ photons set by loss to known dark vibrational states~\cite{vilas2022magneto,Supplemental}. We attribute the additional loss primarily to scattering from the tweezer light during imaging, which we model using the excited electronic state data from above, as described in the Supplemental Material~\cite{Supplemental, holland2023bichromatic}. By comparing the observed number of scattered photons to the calculated loss rates at several wavelengths, we infer that a $\sim$2~ns lifetime of the high-lying electronic states ($\widetilde{D}$, $\widetilde{E}$, $\widetilde{F}$) is consistent with the observed loss~\cite{Supplemental}. This is an order of magnitude shorter than the radiative lifetime of the lower-lying $\widetilde{A}^2\Pi$ and $\widetilde{B}^2\Sigma^+$ states in CaOH, which could indicate non-radiative loss mechanisms, as hypothesized for similar states in tweezer-trapped diatomic molecules~\cite{yu2021coherent,holland2023bichromatic}.

One key feature of the optical tweezer platform is the ability to image molecules nondestructively, for example in order to postselect data on loaded tweezers or to rearrange the molecules into ordered arrays~\cite{endres2016atom}. Doing so requires the ability to distinguish loaded from empty tweezers with high fidelity while minimizing molecule loss. We find that an imaging duration of 7~ms achieves a satisfactory balance between these requirements. Using a background subtraction procedure described in the Supplemental Material~\cite{Supplemental}, we determine that the imaging fidelity after 7~ms of imaging is 95\% for a tweezer loading rate of 30\%, while the probability of the molecule surviving after the image is 80\%. Higher molecule survival probabilities can be attained by imaging for even shorter durations, at the cost of more frequently misidentifying loaded traps as empty. The detection fidelity does not increase significantly for longer imaging durations.

We measure the temperature of the molecules in tweezers after imaging using the release-recapture method (Fig. \ref{fig:3}d)~\cite{tuchendler2008energy}, which we describe here. After loading the tweezer array, we image the molecules for 7~ms to identify loaded tweezers. We then quickly switch off the tweezers, wait a variable time, switch the tweezers back on, and then image a second time (15~ms imaging duration) to determine the fraction of molecules recaptured into the tweezers. The molecule survival probability vs. release time is fit to a Monte Carlo simulation to extract the temperature. For optimal imaging parameters, we measure the molecule temperature to be $T=120(10)$~$\mu$K, corresponding to a ratio of trap depth to molecule temperature of $\eta = 12(1)$. At a sub-optimal single photon detuning $\Delta = 28$~MHz, we instead find that $\eta=9(1)$ but observe no decrease in the maximum number of photons scattered. This suggests that the imaging lifetime is not limited by the molecule temperature.

\section*{Single quantum state preparation and control}

\begin{figure}
    \centering
    \includegraphics{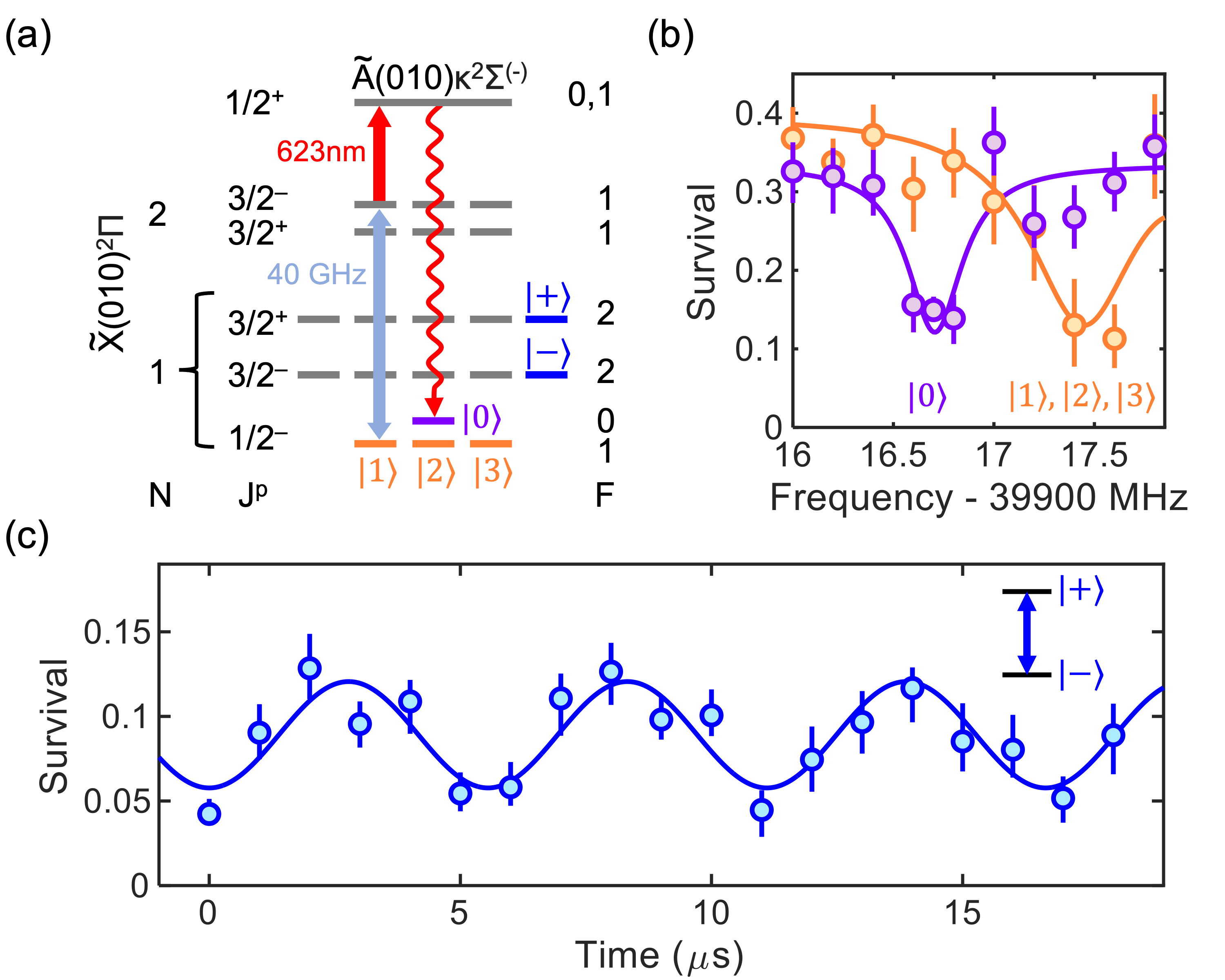}
    \caption{\textbf{Single-state control of CaOH in optical tweezers.} (a) Microwave-optical scheme for pumping population into the single quantum state $|0\rangle$ ($F=0^-$) in the $\widetilde{X}(010)$ bending mode. Microwave radiation at 40~GHz resolves $|0\rangle$ from the nearby states $|1\rangle$, $|2\rangle$, and $|3\rangle$, while a 623~nm laser dissipatively returns population to states $|0\rangle$ through $|3\rangle$. Also shown are the parity-doublet states $|+\rangle$ ($F=2^+,m_F=2$) and $|-\rangle$ ($F=2^-,m_F=2$). Nearby hyperfine levels are omitted from the figure for clarity. (b) Depletion spectroscopy of the $N=1, J=1/2^- \rightarrow N=2, J=3/2^+$ microwave transitions before (orange) and after (purple) optical pumping, demonstrating accumulation of population in $|0\rangle$ (leftmost peak). (c) Rabi oscillations on the $|-\rangle \leftrightarrow |+\rangle$ parity-doublet transition at 43~MHz. Error bars represent 68\% confidence intervals.}
    \label{fig:4}
\end{figure}

Next, we demonstrate coherent control of the internal states of CaOH molecules in the optical tweezer array. We begin by optically transferring molecular population into the $\widetilde{X}^2\Sigma^+(010)$ vibrational bending mode, whose $\ell$-type parity doublet structure is useful for many quantum science applications~\cite{kozyryev2017precision,yu2019scalable,wall2015realizing,anderegg2023quantum,augustovicova2019collisions}. Unlike in our previous work with CaOH in an ODT, where $\sim$1200 photons were scattered during this step~\cite{hallas2023optical}, here with tweezers we optically pump using only $\sim$2 photons by exciting the $\widetilde{X}^2\Sigma^+(000)(N=1^-) \rightarrow \widetilde{A}^2\Pi(010)\kappa^2\Sigma^{(-)}(J=1/2^+)$ transition at 609~nm. The excited state decays to the $N=1,J=1/2^-$ and $N=2,J=3/2^-$ states in the bending mode with approximately equal probability, and the $N=2,J=3/2^-$ population is repumped back through $\widetilde{A}(010)$ and into $N=1$. After optical transfer, the tweezer trap depth is lowered to $\sim$300~$\mu$K in preparation for further state manipulation.

After transfer into the vibrational bending mode, population is predominantly contained in the 4 hyperfine sublevels of the $N=1,J=1/2^-$ manifold, which we label $|0\rangle$, $|1\rangle$, $|2\rangle$, and $|3\rangle$ (Fig.~\ref{fig:4}a). We optically pump the majority of this population into $|0\rangle \equiv (F=0, m_F=0)$ using the microwave-optical pumping scheme outlined in Fig.~\ref{fig:4}a, as demonstrated in previous work with CaOH molecules in an ODT~\cite{anderegg2023quantum}. A microwave field at $\sim$40~GHz resonantly couples states $|1\rangle$, $|2\rangle$, and $|3\rangle$ to $N=2,J=3/2^-$ in the presence of a small electric field (7.5~V/cm), while state $|0\rangle$ is sufficiently detuned (by $\sim$1.4~MHz) to remain dark to the microwaves. A 623~nm laser drives the $N=2,J=3/2^-$ population to $\widetilde{A}(010)\kappa^2\Sigma^{(-)}(J=1/2^+)$, which spontaneously decays to $|0\rangle\ldots|3\rangle$ with $\sim$55\% probability, and back to $N=2,J=3/2^-$ otherwise (with a small $\sim$1\% fraction lost to unaddressed levels). Population therefore accumulates in $|0\rangle$ after approximately 7 photons are scattered. Finally, we perform microwave spectroscopy to read out the hyperfine state populations~\cite{Supplemental}. As shown in Fig.~\ref{fig:4}b, we find that the molecular population is predominantly contained in the near-degenerate $F=1^-$ states ($|1\rangle$, $|2\rangle$, and $|3\rangle$) in the absence of optical pumping (orange data), while a majority of the molecular population is transferred into $|0\rangle$ after optical pumping is applied (purple data).

After optical pumping, we demonstrate coherent control of CaOH molecules in the optical tweezer array by driving Rabi oscillations on a parity-doublet transition in the bending mode. We choose the hyperfine stretched states $|+\rangle \equiv (F=2^+,m_F=2)$ and $|-\rangle \equiv (F=2^-,m_F=2)$ due to their large transition dipole moment, which would enable strong dipolar spin-exchange interactions between adjacent molecules and make them promising qubit states. We begin by transferring population from $|0\rangle$ to $|+\rangle$ with an rf $\pi$ pulse in the presence of a small magnetic field $B=2$~G. We then resonantly drive the $|+\rangle \leftrightarrow |-\rangle$ parity-doublet transition with a 43~MHz rf field. The population detected in $|-\rangle$ as a function of drive time is shown in Fig.~\ref{fig:4}c, demonstrating clear Rabi oscillations with a frequency $\Omega = 2\pi \times 180$~kHz.

\section*{Summary and Outlook}

In summary, we have realized an array of single CaOH molecules by directly loading laser-cooled molecules into tightly focused optical tweezer traps at 785~nm. We directly and nondestructively image individual trapped molecules and achieve fidelities greater than 90\% for distinguishing loaded from empty traps. We have characterized the trap wavelength dependence of the tweezer imaging, which is found to be limited by tweezer light-mediated coupling to high-lying electronic states in CaOH. Finally, we prepared CaOH molecules in the tweezer array in single quantum states and observed coherent Rabi oscillations on a parity-doublet transition in the vibrational bending mode.

These results pave the way for numerous experiments in quantum information science~\cite{wei2011entanglement,yu2019scalable}, quantum simulation~\cite{wall2013simulating,wall2015realizing}, ultracold collisions~\cite{cheuk_2020,anderegg_2021,augustovicova2019collisions}, and precision measurements~\cite{kozyryev2017precision,anderegg2023quantum,kozyryev2021enhanced} that require (or could benefit from) the combination of full internal and external state control of individual polyatomic molecules, as provided by the optical tweezer platform. The platform demonstrated here can be used in the near term to realize dipolar interactions between molecules in adjacent optical tweezers in parity doublet states and for cooling of the molecules' motion to the quantum regime~\cite{lu2023raman,bao2023raman}. The imaging fidelities achieved here could be increased in future work by using longer tweezer light wavelengths, e.g. 1064~nm, which we successfully used to optically trap CaOH in larger optical dipole traps~\cite{hallas2023optical}. Finally, this work is expected to be extendable to larger and/or more complex polyatomic molecules amenable to direct laser cooling techniques, including symmetric and asymmetric top molecules~\cite{mitra2020direct, augenbraun2020molecular}.

We thank L. Cheng for providing dynamic polarizability calculations. This work was supported by the AFOSR, NSF, ARO, and DOE Quantum Systems Accelerator. PR acknowledges support from the NSF GRFP, and GKL and LA acknowledge support from the HQI.

\bibliographystyle{apsrev4-2}
\bibliography{CaOHReferences}

\pagebreak

\clearpage

\noindent \Large\textbf{Supplemental Material}
\bigskip

\renewcommand{\thefigure}{S\arabic{figure}}
\setcounter{figure}{0}

\normalsize
\noindent\textbf{Experimental details}
\newline
The beginning of our experimental sequence is similar to that used in previous work~\cite{vilas2022magneto,hallas2023optical,anderegg2023quantum}, with small modifications. CaOH molecules are produced in a cryogenic buffer-gas beam, radiatively slowed using the frequency-chirped slowing method~\cite{andereggthesis,truppe2017intense}, and trapped in a magneto-optical trap (MOT)~\cite{vilas2022magneto}. The vibrational repumping scheme is the same as described in Ref. \cite{vilas2022magneto}, except that an additional repumping laser addressing the $\widetilde{X}^2\Sigma^+(03^10)(N''=1) \rightarrow \widetilde{A}(010)\mu^2\Sigma^{(+)}(J'=1/2^+)$ transition at 655~nm has been added. This increases the vibrational-decay-limited average number of photons scattered per molecule from $\sim$12000 to $\sim$16000.

The optical dipole trap is generated from a 13.3~W, 1064~nm laser beam, focused down to a $\sim$25~$\mu$m waist in the center of the molecular MOT~\cite{hallas2023optical}. Rastering of the optical dipole trap (ODT) position during tweezer loading is achieved with a piezo-actuated mirror mount in the 1064~nm beam path. The tweezer trap array is generated by a 784.5~nm laser beam which is diffracted by an acousto-optic deflector (AOD; AA Opto Electronic DTSX-400-780) in the Fourier plane of the imaging system to generate the individual traps. The AOD is driven with the sum of 6 rf tones centered near 100~MHz and spaced by 1.875~MHz. The relative phases of the rf tones are chosen using a random sampling procedure to minimize the peak amplitude of the summed waveform. The resulting beams are expanded and then focused through an in-vacuum aspheric lens (Edmund Optics, 25mm diameter, 0.40 NA) to generate an array of 6 tweezer traps with $\sim$2~$\mu$m waists, separated by 11~$\mu$m. The tweezer light is co-linear with but counterpropagating to the ODT beam.

In order to minimize background scattered light during imaging of the tweezer array, six small 1.5~mm $1/e^2$ diameter beams are used for the $\Lambda$ imaging. These are installed at a small angle to the large 10~mm $1/e^2$ diameter beams used for the MOT and ODT/tweezer loading, which are turned off during tweezer array imaging. Additionally, the repumping laser beam size is reduced to $\sim$5~mm in diameter after radiative slowing, MOT, and ODT/tweezer loading by closing a mechanical shutter with a small aperture drilled in the center. This is necessary because several of the repumping laser wavelengths are $<$5~nm away from the 626~nm tweezer fluorescence and are inefficiently filtered from the imaging system.

The optimal parameters for tweezer imaging are as follows (detunings are referenced relative to free space resonance). The single-photon detuning is $\Delta = 22$~MHz, the two-photon detuning is $\delta = 1.25$~MHz, the sideband intensity ratio is $I_{1/2}/I_{3/2}\approx2$, and the total intensity is approximately $9$~mW cm$^{-2}$ per beam. The tweezer trap depth used for imaging is 1.4~mK. Additionally, four of the repumping laser frequencies are tuned from the free-space optimal used for radiative slowing, the MOT, and tweezer loading, to AC Stark-shifted frequencies optimized for tweezer imaging brightness. Specifically, the $\widetilde{X}^2\Sigma^+(100)\rightarrow\widetilde{B}^2\Sigma^+(000)$ repumping laser is tuned approximately 60~MHz to the blue; the $\widetilde{X}^2\Sigma^+(010)(N=1)\rightarrow\widetilde{B}^2\Sigma^+(000)$ repumping laser is tuned approximately 140~MHz to the blue; the $\widetilde{X}^2\Sigma^+(02^00)\rightarrow\widetilde{A}^2\Pi_{1/2}(100)$ repumping laser is tuned approximately 40~MHz to the red; and the $\widetilde{X}^2\Sigma^+(010)(N=2)\rightarrow\widetilde{A}(010)\kappa^2\Sigma^{(-)}$ repumping laser is tuned approximately 140~MHz to the blue.
Note that the frequency spectrum of each repumping laser is broadened by $\sim$300~MHz to the red of free-space resonance for radiative slowing~\cite{vilas2022magneto}. This broadening remains on throughout the experimental sequence, therefore reducing the precision requirements for the tweezer repumping frequencies.

Microwave-optical pumping of CaOH into a single quantum state is achieved using 40~GHz microwave radiation introduced via a horn antenna mounted outside the vacuum chamber. The optical pumping light comes from the $\widetilde{X}^2\Sigma^+(010)(N=2)\rightarrow \widetilde{A}(010)\kappa^2\Sigma^{(-)}$ repumping laser. Radio-frequency fields for the coherent state manipulation and Rabi oscillations are introduced by directly driving the in-vacuum rf MOT coils with an oscillating voltage.

Spectroscopy used to determine the population in states $|0\rangle \cdots |3\rangle$ before and after optical pumping (Fig.~4b of the main text) is performed by applying microwave radiation near the $N=1, J=1/2^-, F=0,1 \rightarrow N=2, J=3/2^+, F=1,2$ transitions around 39917~MHz. Population excited to $N=2$ is depleted with a 627~nm laser addressing the $\widetilde{X}^2\Sigma^+(010)(N=2, J=3/2^+)\rightarrow \widetilde{A}(010)\mu^2\Sigma^{(+)}(J=5/2^-)$ transition, which decays to rotational states that are dark to the tweezer imaging light. The spectroscopy data is acquired by imaging the tweezer array for 7~ms immediately after loading, then proceeding to transfer molecules into the bending mode, optically pump, and then perform microwave spectroscopy. The molecules are then imaged for 15~ms. The plotted molecule survival is the fraction of molecules observed in the final image, conditioned on their presence in the first image. A similar procedure is used for the Rabi oscillation data in Fig.~4c, with the molecule survival determined using an initial 7~ms image before optical pumping and coherent control.

\bigskip
\noindent\textbf{Imaging system}
\newline
Fluorescence at 626~nm from the tweezer-trapped molecules is collected with the in-vacuum aspheric lens and imaged onto an EMCCD camera (Andor iXon Ultra 897). Spherical abberation correction plates (3$\times$ $-1.00\lambda$; Edmund Optics) are used to compensate abberations arising from the chromatic focal shift between the 785~nm tweezers and the 626~nm fluorescence. The collection efficiency is expected to be approximately 1\%, limited by the effective numerical aperture of the imaging optics ($\sim$3\% collection efficiency) and by transmission losses from the optical elements in the imaging path ($\sim$40\%). In practice, we find by comparing the observed histograms to simulations that the collection is slightly lower, likely due to imperfect alignment. Images are acquired by binning the fluorescence from each tweezer onto a single EMCCD pixel at the hardware level, minimizing the effect of camera readout noise. This procedure applies to all data presented in the main text except for the averaged image in Fig.~1a, which uses no binning.

\bigskip
\noindent\textbf{Tweezer trap frequencies and trap depth}
\newline
The oscillation frequencies of molecules in the optical tweezer array are measured using the parametric heating method. After loading the array, molecules are imaged for 7~ms to detect loaded traps. The trap intensity is then modulated at a variable frequency $\omega_\text{mod}$ for 5~ms, causing parametric heating loss when $\omega_\text{mod}\approx 2\omega_i$, where $\omega_{i=x,y,z}$ are the trap frequencies. The tweezers are briefly shut off for 6~$\mu$s to allow heated molecules to escape, then the remaining molecules are imaged for 15~ms. The molecule survival exhibits loss features at $\omega_\text{mod}=2\pi\times 119(4)$~kHz and $\omega_\text{mod}=2\pi\times 167(5)$~kHz, corresponding to radial trap frequencies of $\omega_x = 2\pi \times 59(2)$~kHz and $\omega_y = 2\pi \times 84(2)$~kHz.
Using the calculated 780~nm dynamic polarizability for the ground state of CaOH ($\alpha_0 = 324.3$~a.u.~\cite{780polarizability}) and the known laser power per trap ($\sim$125~mW), these frequencies imply a tweezer trap depth of $U_0 = k_B \times 1.4$~mK, an axial frequency of $\omega_z \approx 7$~kHz, and slightly elliptical tweezer beams with waists $w_x \approx 2.4$~$\mu$m and $w_y \approx 1.7$~$\mu$m.

\bigskip
\noindent\textbf{Calibration of number of photons scattered}
\newline
The number of photons scattered during tweezer imaging is calibrated by comparing the collected fluorescence to reference images taken with a known number of photons limited by vibrational repumping. Specifically, to calibrate the maximum number of photons that can be scattered, we take 200~ms images in two configurations: (1) all repumping lasers on (fluorescence level $S_1$), and (2) only the $\widetilde{X}(100)$, $\widetilde{X}(02^00)$, $\widetilde{X}(200)$, $\widetilde{X}(12^20)$, and $\widetilde{X}(110)$ repumping lasers on, limiting the average number of photons scattered to 460 (fluorescence level $S_2$)~\cite{vilas2022magneto, zhang2021accurate}. The total number of photons scattered before molecules are lost during tweezer imaging with all repumping lasers is therefore $N_\text{im} = S_1/S_2 \times 460$. For the optimal imaging parameters used in this work, we find $N_\text{im} = 8(1)\times 10^3$.

\bigskip
\noindent\textbf{Modeling excited state AC Stark shifts}
\newline
The tweezer light is near-resonant with transitions from $\widetilde{A}^2\Pi$ and $\widetilde{B}^2\Sigma^+$ state vibrational levels used for laser cooling and repumping to vibrational levels of the high-lying $\widetilde{D}^2\Sigma^+$, $\widetilde{E}^2\Sigma^+$, and $\widetilde{F}$ electronic states~\cite{pereira1996observation}. In the large detuning limit ($\Delta \gg \Omega$, where $\Delta$ is the tweezer detuning and $\Omega$ is the Rabi frequency), the effect of these couplings on the energy of excited state $i$ is given by
\begin{equation}
\Delta E_i/\hbar \approx \sum_j \frac{\Omega_{ij}^2}{4(\omega_{ij} - \omega_L)} + \frac{\Omega_{ij}^2}{4(\omega_{ij} + \omega_L)}
\label{eq:ACStarkFull}
\end{equation}
where $\Omega_{ij}$ is the Rabi frequency for the transition from state $i$ to state $j$, $\omega_L$ is the laser frequency, $\omega_{ij} = (E_j-E_i)/\hbar$ is the transition frequency, and the sum is over all molecular states $j$ that can couple to state $i$.

If we restrict ourselves to a small range of wavelengths $\Delta \lambda$ near the tweezer wavelength $\lambda_0$, we can make the approximation that only a small number of near-detuned states $k$ are responsible for the wavelength dependence, and that the remaining states contribute an approximately constant energy offset $\Delta E_\text{off}/\hbar$ as long as they are far detuned compared to $\Delta \lambda$. Furthermore, we can make the rotating wave approximation and drop the second, counter-rotating term in eqn.~\ref{eq:ACStarkFull}. We find that the wavelength dependence of the energy of state $i$ in the range $\Delta \lambda$, which contains $N$ near-detuned states $k$, is
\begin{equation}
\Delta E_i/\hbar \approx \sum_{k=1}^N \frac{\Omega_{ik}^2}{8\pi c (1/\lambda_L - 1/\lambda_{ik})} + \Delta E_{\text{off},i}/\hbar
\label{eqn:ACStarkLambda}
\end{equation}
where $\lambda_L = 2\pi c/\omega_L$ and $\lambda_{ik} = 2\pi c(1/\omega_i - 1/\omega_k)$. This expression is used to fit the AC Stark shift data in Fig.~2b of the main text, where the Rabi frequencies $\Omega_{ik}$, center wavelengths $\lambda_{ik}$, and energy offsets $\Delta E_{\text{off},i}$ are fit as free parameters.

\bigskip
\noindent\textbf{Measured AC Stark shifts}
\newline
The excited state AC Stark shifts are measured by imaging molecules in the tweezer array with one of the repumping lasers replaced by a separate frequency-tunable beam. At each tweezer wavelength, the frequency of the repumping beam is scanned near resonance, and the AC Stark shift is determined by comparing the repumping frequency that maximizes the image brightness with the known free space resonance for the repumping transition. The specific transitions used are $\widetilde{X}^2\Sigma^+(100)\rightarrow\widetilde{B}^2\Sigma^+(000)$, $\widetilde{X}^2\Sigma^+(02^00)\rightarrow\widetilde{A}^2\Pi_{1/2}(100)$, and $\widetilde{X}^2\Sigma^+(010)(N=2)\rightarrow\widetilde{A}(010)\kappa^2\Sigma^{(-)}$. For the $\widetilde{A}^2\Pi_{1/2}(000)$ state AC Stark shifts, the $\widetilde{X}^2\Sigma^+(000)\rightarrow\widetilde{A}^2\Pi_{1/2}(000)$ cooling light was instead tuned during tweezer imaging, and the AC Stark shift was determined by comparing the optimal tweezer imaging detuning with the optimal free space $\Lambda$ imaging detuning. This approach is susceptible to common mode offsets in the measured AC Stark shift due to the need to compare with a free space resonance frequency. However, relative frequencies and resonance wavelengths are insensitive to such offsets, meaning that the extracted transition wavelengths and transition strengths (Rabi frequencies) are robustly determined, in turn leading to accurate loss rate calculations.

The measured transition wavelengths determined from the fit to eqn.~\ref{eqn:ACStarkLambda} (Fig.~2b of the main text) are as follows.
The energy of the $\widetilde{A}^2\Pi_{1/2}(000)$ excited state exhibits a resonance at $777.9(6)$~nm, which is consistent with the expected wavelength for coupling from $\widetilde{A}^2\Pi_{1/2}(000)\rightarrow \widetilde{D}^2\Sigma^+(100)$~\cite{pereira1996observation}.
$\widetilde{A}^2\Pi_{1/2}(100)$ contains resonances at $777$~nm and $789$~nm (consistent with coupling to $\widetilde{D}^2\Sigma^+(200)$ and $\widetilde{D}^2\Sigma^+(110)$, respectively), while $\widetilde{A}(010)\kappa^2\Sigma^{(-)}$ displays a strong resonance at $780.6$~nm due to coupling with the $\widetilde{D}^2\Sigma^+(110)$ state. Three features are observed in the $\widetilde{B}^2\Sigma^+(000)$ data, at $782.4(1)$~nm, $783.1(1)$~nm, and $787.2(1)$~nm, corresponding to high-lying features at 30804(2)~cm$^{-1}$, 30792(2)~cm$^{-1}$, and 30726(2)~cm$^{-1}$, respectively. These features fall slightly above the origin energy of the $\widetilde{E}^2\Sigma^+$ and $\widetilde{F}$ electronic potentials, in a frequency range not covered by previous spectroscopic literature~\cite{pereira1996observation}.

\bigskip
\noindent\textbf{Modeling losses due to excited state coupling}
\newline
Losses arising from coupling of state $i$ (in the laser cooling scheme) to a lossy excited state $j$ occur at a rate $R_{ij} = |c_{ij}|^2\Gamma_j \rho_{ii}$, where $c_{ij}$ is the admixture of state $j$ into $i$ due to tweezer coupling, $\Gamma_j$ is the decay rate of state $j$ (assumed to be purely to nondetectable states), and $\rho_{ii}$ is the molecular population in state $i$ during tweezer imaging~\cite{holland2023bichromatic}. In the large detuning limit $\Delta_{ij} \gg \Omega_{ij}$, the admixture is $c_{ij} \approx \Omega_{ij}/(2\Delta_{ij})$, where $\Delta_{ij} = \omega_L - \omega_{ij}$. The population of the laser cooling excited state $i$ is approximately $\rho_{ii} \approx \Gamma_\text{scatt}/\Gamma_i \times r_i$, where $\Gamma_\text{scatt}=250\times10^3$~s$^{-1}$ is the photon scattering rate during imaging and $r_i$ is the fraction of photon scattering events that originate from state $i$. The latter quantity is set by the vibrational state repumping scheme used in the experiment~\cite{vilas2022magneto}.

The total observed molecule loss rate during tweezer imaging is a combination of excited state loss $R_{ij}$, vacuum loss $R_\text{vac}=1/690$~ms$^{-1}$ (see main text), and loss to dark (i.e., not repumped) vibrational states, $R_\text{vib} = \Gamma_\text{scatt}/N_\text{phot}$, where $N_\text{phot} = 1.6\times10^4$ is the vibrational repumping-limited average number of photons scattered per molecule. The total loss rate $R_\text{loss} = R_{ij} + R_\text{vib} + R_\text{vac}$ is calculated as a function of tweezer wavelength using the fitted Rabi frequencies and center wavelengths from the AC Stark shift data.

\begin{figure}
    \centering
    \includegraphics[]{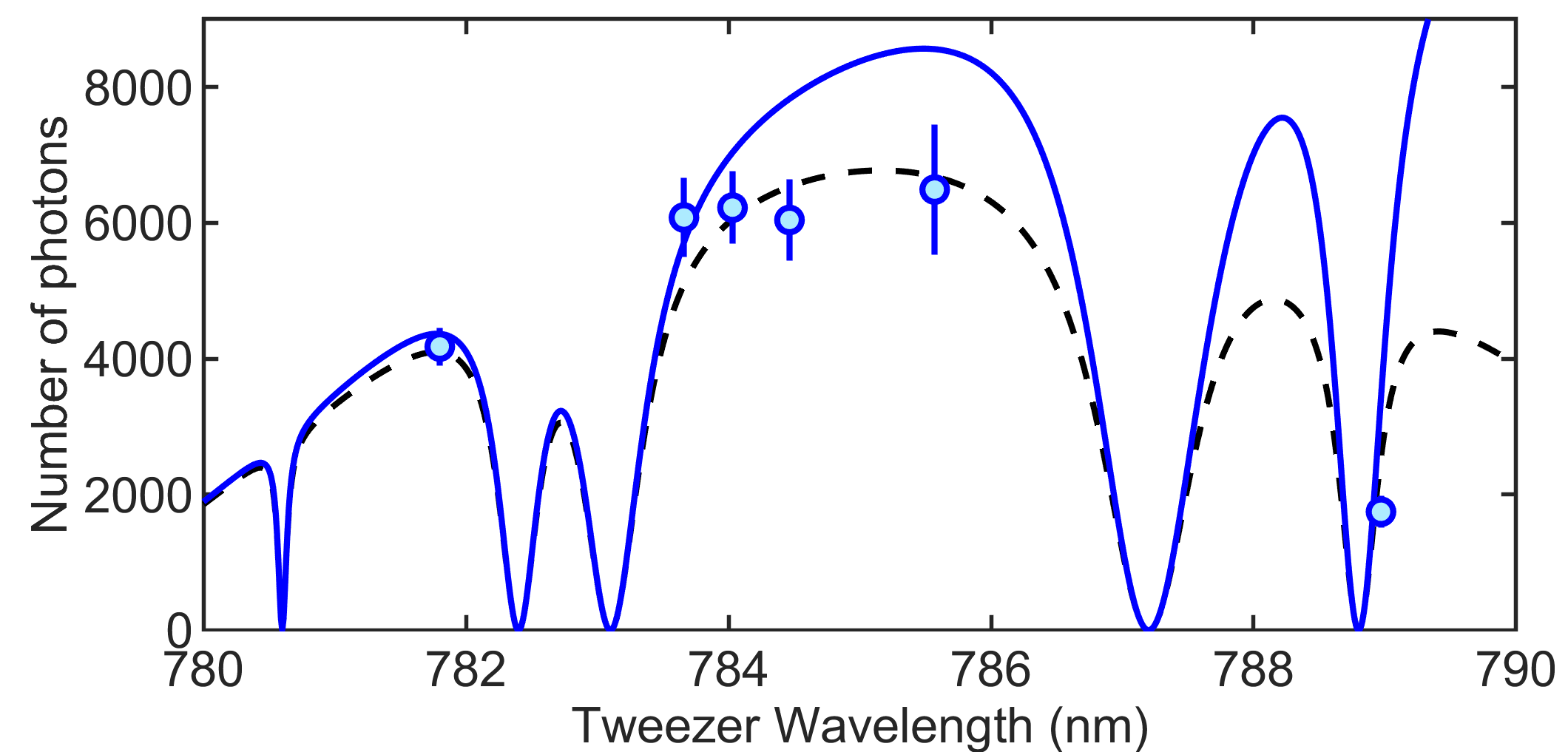}
    \caption{\textbf{Imaging photon budget vs. tweezer wavelength.} Calculated (curves) and measured (points) number of imaging photons that can be scattered during molecule imaging in the tweezer array, as a function of the trapping wavelength. Calculations include the effect of vacuum loss, loss to vibrational dark states, and trap-wavelength-dependent excitation to lossy excited electronic levels. The solid blue curve accounts for excitation only to those states observed using AC Stark shift data, while the black, dashed curve additionally includes $\widetilde{A}^2\Pi_{1/2}(000)\leftrightarrow \widetilde{D}^2\Sigma^+(010)$ excitation predicted at 793.6~nm.}
    \label{fig:PhotonsVsWavelength}
\end{figure}

Fig~\ref{fig:PhotonsVsWavelength} shows the number of photons scattered per molecule in the experiment (measured as described above) as a function of tweezer wavelength.
We also plot the calculated photon number determined from the loss rate calculations, $N_\text{im} = \Gamma_\text{scatt}/R_\text{loss}$ (blue, solid curve). Here we have used the known loss rates $R_\text{vac}$ and $R_\text{vib}$ as well as the measured Rabi frequencies and wavelengths for tweezer-induced excited-state loss. The only unknown parameter is $\Gamma_j$, the loss rate from the high-lying states to which the tweezer couples, which was assumed to be constant for all excited states and varied to match the measured photon number data. We find reasonable agreement with the experimental data using an excited state loss rate $\Gamma_j \approx 2\pi \times 75$~MHz, corresponding to a lifetime of $\Gamma_j^{-1} \approx 2.1$~ns. The remaining disagreement, particularly in the 784-786~nm range, may be attributed to the effect of other excited state resonances that were not considered in the AC Stark shift data. For example, there is a known resonance from $\widetilde{A}^2\Pi_{1/2}(000)\leftrightarrow \widetilde{D}^2\Sigma^+(010)$ at 793.6~nm~\cite{pereira1996observation}. Including this state, assuming the same Rabi frequency measured for the $\widetilde{A}^2\Pi_{1/2}(000)\leftrightarrow \widetilde{D}^2\Sigma^+(100)$ transition at 777.9~nm, results in better agreement with the experimental data (black, dashed curve).

\bigskip
\noindent\textbf{Imaging fidelity}
\newline
The imaging fidelity is defined as the probability of correctly identifying whether a trap is loaded (1 molecule) or empty (0 molecules). Specifically, it may be defined in terms of the error rates $\epsilon_{01}$ (incorrectly identifying an empty trap as loaded; ``false positive") and $\epsilon_{10}$ (incorrectly identifying a loaded trap as empty; ``false negative") as follows: $f(p) = 1 - p\epsilon_{10} - (1-p)\epsilon_{01}$, where $p$ is the tweezer loading probability~\cite{holland2023bichromatic}.

\begin{figure*}
\centering
\includegraphics[]{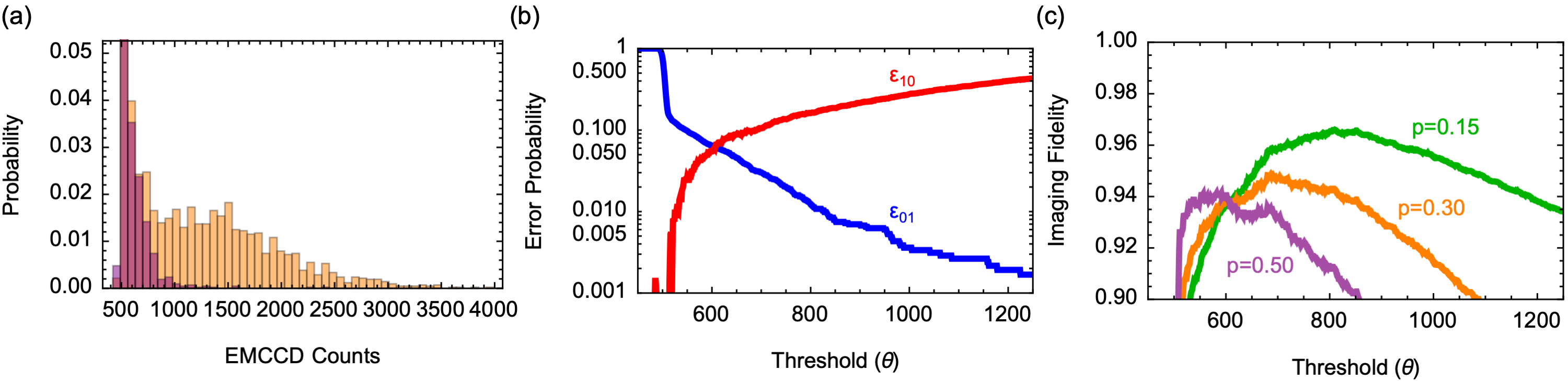}
\caption{\textbf{Determination of tweezer imaging fidelities.} (a) Signal (orange) and background (purple) histogram data, $h_\text{exp}(n)$ and $h_0(n)$, for 7~ms tweezer images. The average tweezer loading probability in the signal histogram is $p=0.34$. (b) Misidentification error rates $\epsilon_{01}(\theta)$ and $\epsilon_{10}(\theta)$ for 7~ms tweezer images, inferred from the experimental data as described in the text. (c) Imaging fidelities $f(p)$ for several average tweezer loading probabilities $p$, as described in the text.}
\label{fig:ImagingFidelity}
\end{figure*}

The error rates depend on the threshold number of counts, $\theta$, used to distinguish loaded and empty traps. The false positive rate, $\epsilon_{01}(\theta)$, is determined by analyzing histogram data for empty traps, which can be collected by taking images of tweezers guaranteed to be empty by, e.g., turning off the molecule ablation source. The histogram data, $h_0(n)$, is a function of the number of image counts $n$, and is normalized so that $\sum_n h_0(n) = 1$. Moreover, we define a cumulative probability distribution for the histogram data, $g_0(\theta) = \sum_{n=0}^\theta h_0(n)$. Following these definitions, the false positive rate $\epsilon_{01}(\theta)$ is simply the probability of collecting greater than $\theta$ counts from an image of an empty trap, i.e. $\epsilon_{01}(\theta) = 1-g_0(\theta)$.

To determine the false negative rate $\epsilon_{10}(\theta)$, we must infer the histogram of image counts collected from loaded traps, $h_1(n)$. Because the loading rate is much less than 1 in the experiment, this requires subtracting out the effect of empty traps from $h_\text{exp}(n)$, the measured histogram data. This is done by scaling the empty trap histogram data $h_0(n)$ by a factor of $(1-p)$ (the probability of a trap being empty), subtracting this from the experimental data to remove the background contribution, and renormalizing the result: $h_1(n) = \left[h_\text{exp}(n) - (1-p)h_0(n)\right]p^{-1}$. We discuss how $p$ is determined below. The cumulative probability distribution for loaded traps is defined as $g_1(\theta) = \sum_{n=0}^\theta h_1(n)$, and the false negative error rate is the probability of collecting fewer than $\theta$ counts from a loaded tweezer, $\epsilon_{10}(\theta) = g_1(\theta)$.

Determining $p$, the tweezer loading probability, is equivalent to finding the scale factor to apply to $h_0(n)$ so that the empty trap background is appropriately subtracted from $h_\text{exp}(n)$ in the procedure above. To perform this background subtraction, we take advantage of the tall ($\sim$80\% of the counts in $h_0(n)$), zero-photoelectron peak generated by the EMCCD camera, which is approximately Gaussian in shape with a standard deviation of $\sigma_r\approx4$ digital counts (given by the camera readout noise), and is centered at zero photoelectron counts (far left side of the histogram). In order to perform the background subtraction, we assume that this zero-photoelectron peak in the experimental data arises purely from empty traps, meaning that we can rescale the background histogram to match the height of this peak and assume that the result accurately reflects the distribution of empty trap counts in the experimental data. This approximation is justified by the fact that the \emph{average} number of counts arising from loaded traps, $\bar{n}$, is hundreds of counts above this zero-photon peak ($\bar{n} \gg \sigma_r$), so that any residual tail at low count numbers is negligible compared to the contribution from empty traps.

To determine $p$, we isolate the zero-photon peak in both the signal and background histograms, then apply a scale factor of $(1-p)$ to the empty histogram data and calculate the sum of squared differences between the two, $S(p) = \sum_{n=n_i}^{n_f} \left[h_\text{exp}(n) - (1-p)h_0(n) \right]^2$, where $n_i$ and $n_f$ are the lower and upper edges of the peak. The value of $p$ that minimizes $S(p)$ is the inferred loading probability for the experimental data and is used to determine $h_1(n)$ via the procedure defined above.

Fig.~\ref{fig:ImagingFidelity} illustrates the fidelity results for 7~ms tweezer images, chosen to be short enough for reasonably high molecule survival after imaging. Fig.~\ref{fig:ImagingFidelity}a shows the signal and background histograms ($h_\text{exp}(n)$ and $h_0(n)$, respectively) from which the error probabilities $\epsilon_{01}(\theta)$ and $\epsilon_{10}(\theta)$ are inferred using the procedure described above. These error rates are plotted as a function of the threshold value $\theta$ in Fig.~\ref{fig:ImagingFidelity}b. Higher choices of threshold reduce the probability of misidentifying empty traps as loaded ($\epsilon_{01}$, blue curve), at the expense of increasing the likelihood of misidentifying loaded traps as empty ($\epsilon_{10}$, red curve). Fig.~\ref{fig:ImagingFidelity}c shows the imaging fidelity, $f(p,\theta) = 1 - p\epsilon_{10}(\theta) - (1-p)\epsilon_{01}(\theta)$, as a function of detection threshold for various loading probabilities. Each curve exhibits a peak corresponding to the optimal tradeoff between the two misidentification error sources. For higher loading probabilities, a lower threshold optimizes the fidelity, since reducing the likelihood of misidentifying loaded traps as empty becomes more important at higher loading rates. For all loading rates, fidelities $f \gtrsim 0.94$ are possible for an appropriate choice of threshold.

\end{document}